\documentclass{openjournal}
\pdfoutput=1

\usepackage{graphicx}
\graphicspath{ {paper2_extended/} }
\usepackage{amssymb,amsmath,bm}
\usepackage{pxfonts}
\usepackage{aas_macros}

\newcommand{\nv}{\hat{\bm n}}

\newcommand{\deq}{\coloneqq}

\newcommand{\Ecut}{E_\text{cut}}
\newcommand{\Ecuti}[1]{E_{\text{cut},\,#1}}
\newcommand{\Emax}{E_\text{max}}
\newcommand{\cN}{\mathcal{N}}
\newcommand{\cS}{\mathcal{S}}

\newcommand{\de}{\mathrm{d}}

\newcommand{\ra}{\rightarrow}

\newcommand{\fpr}{f_\text{p}}

\newcommand{\nev}{N_\text{ev}}
\newcommand{\snr}{\text{SNR}}
\newcommand{\fsky}{\ensuremath{f_\text{sky}}}
\newcommand{\bcut}{b_\text{cut}}

\def\dd{ \text{d} }

\def\+{\dagger}
\def\la{\langle}
\def\ra{\rangle}

\def\n0{{\la n \ra}}

\def\B0{{\la B \ra}}

\let\Gamma\varGamma
\let\Delta\varDelta
\let\Theta\varTheta

\let\Lambda\varLambda
\let\varLambda\oldLambda
\let\Xi\varXi
\let\Pi\varPi
\let\Sigma\varSigma
\let\Upsilon\varUpsilon
\let\Phi\varPhi
\let\Psi\varPsi
\let\Omega\varOmega

\begin{document}

\shorttitle{Modelling UHECR-galaxies correlations}
\title{Modelling cross-correlations of ultra-high-energy cosmic rays and galaxies}
\author{Federico R.~Urban$^{a}$}
\author{Stefano Camera$^{b,\,c}$}
\author{and David Alonso$^{d}$}

\affiliation{$^{a}$CEICO, FZU -- Institute of Physics of the Czech Academy of Sciences,\\Na Slovance 1999/2, 182 21 Prague, Czech Republic}
\affiliation{$^{b}$Dipartimento di Fisica, Universit\`a degli Studi di Torino,\\Via P.\ Giuria 1, 10125 Torino, Italy}
\affiliation{$^{c}$INFN -- Istituto Nazionale di Fisica Nucleare, Sezione di Torino,\\Via P.\ Giuria 1, 10125 Torino, Italy}
\affiliation{$^{d}$Department of Physics, University of Oxford,\\Denys Wilkinson Building, Keble Road, Oxford OX1 3RH, UK}

\begin{abstract}
The astrophysical engines that power ultra-high-energy cosmic rays (UHECRs) remain to date unknown. Since the propagation horizon of UHECRs is limited to the local, anisotropic Universe, the distribution of UHECR arrival directions should be anisotropic. In this paper we expand the analysis of the potential for the angular, harmonic cross-correlation between UHECRs and galaxies to detect such anisotropies. We do so by studying simulations performed assuming proton, oxygen and silicon injection models, each simulation containing a number of events comparable to a conservative estimate of currently available datasets, as well as by extending the analytic treatment of the magnetic deflections. Quantitatively, we find that, while the correlations for each given multipole are generally weak, (1) the total harmonic power summed over multipoles is detectable with signal-to-noise ratios well above~5 for both the auto-correlation and the cross-correlation (once optimal weights are applied) in most cases studied here, with peaks of signal-to-noise ratio around between~8 and~10 at the highest energies; (2) if we combine the UHECR auto-correlation and the cross-correlation we are able to reach detection levels of \(3\sigma\) and above for individual multipoles at the largest scales, especially for heavy composition. In particular, we predict that the combined-analysis quadrupole could be detected already with existing data.
\end{abstract}

\maketitle

\section{Introduction}\label{sec:intro}

Sixty-two years ago the first ultra-high-energy cosmic ray (UHECR) was detected \citep{Linsley:1961kt}; in the decades since observations have continued to accumulate UHECRs up to extreme energies exceeding \(100\,\mathrm{EeV}\) (\(10^{20}\,\mathrm{eV}\)). The mystery of how to accelerate particles, which appear to be overwhelmingly charged nuclei, to such extraordinary energies remains as elusive as ever \citep{AlvesBatista:2019tlv}. There is no compelling theoretical model for the UHECR acceleration mechanism among several promising candidates. More importantly, despite tremendous recent experimental progress \citep{TelescopeArray:2021dfb,TelescopeArray:2021gxg,TelescopeArray:2021ygq,Kim:2021mcf,TelescopeArray:2021dpk,TelescopeArray:2021dzr,PierreAuger:2021dqp,PierreAuger:2021rfz,Mayotte:2021Wq,TelescopeArray:2023waz}, we have not yet been able to trace the UHECRs back to their sources. The difficulty in identifying the sources of UHECRs from the data stems from two coupled problems: it is very hard to measure the charge, or atomic number \(Z\), of UHECRs with good accuracy \citep{Zhezher:2021qke,PierreAuger:2021mmt} and it is very hard to realistically model the three-dimensional structure of the Galactic magnetic field (GMF) from integrated two-dimensional information \citep{Boulanger:2018zrk}; moreover, little is known about possible extra-Galactic magnetic fields, which, if present, also further contribute to UHECR deflections away from their sources.\footnote{There is an additional problem: the accuracy and precision with which the energy of a UHECR event is determined has a systematic uncertainty of approximately 15\% to 20\% \citep{TelescopeArray:2021zox} as well as a resolution of the order of 10\% \citep{Abbasi:2023swr,Fenu:2023men}. This is, however, not as relevant as the uncertainties on UHECR charge and intervening magnetic fields.} Thus, the arrival directions of UHECRs as observed from Earth are displaced from their arrival directions before encountering the GMF distorting ``lens'' by a mostly unknown angle \(\theta\propto ZB/E\), where \(B\) is the GMF strength, in a mostly unknown direction.

In order to try to bypass some of these issues and to extract as much information as possible from the sky distribution of arrival directions of UHECRs, in particular from its anisotropies, in a previous work we have proposed a new observable, the angular, harmonic cross-correlation (XC) between the angular distribution of galaxies and that of UHECR arrival directions, and assessed its performance in a proton-only injection model \citep{Urban:2020szk}. In that work we showed how the XC is a complementary observable to the more familiar angular, harmonic UHECR auto-correlation (AC): (i) the XC is more sensitive to small-scale angular anisotropies and in certain conditions can be easier to detect than the AC---conversely, the AC is more sensitive to large-scale anisotropies; (ii) because systematic uncertainties do not cross-correlate and, under some conditions, statistical noises do not strongly cross-correlate either, the XC is an experimentally cleaner observable; (iii) the XC encodes more astrophysical information than the AC, and, since the two respond differently to model parameters, taken together they can break degeneracies. The AC and XC have been already recognised as useful tools to detect the angular anisotropy of the UHECR flux \citep{Sommers:2000us,Tinyakov:2014fwa,Deligny:icrc2015,Ahlers:2017wpb,diMatteo:2018vmr,Tanidis:2022jox,Tanidis:2023pby}.

In this paper we extend and improve our analysis in several ways, yet keeping with a mostly analytic treatment. (1) Beyond the pure hydrogen model (\({^{1}}\)H) with injection slope \(\gamma=2.6\), we extend the analysis to an oxygen model (\({^{16}}\)O) with injection slope \(\gamma=2.1\) at all energies, and a silicon (\({^{28}}\)Si) model with \(\gamma=1.5\) and a sharp \(\Emax=280\,\mathrm{EeV}\) energy cutoff, inspired by the analysis of \cite{diMatteo:2017dtg}. (2) We extend the analysis to different number of detected events to forecast the progress that can be made with future observations. (3) We develop a more realistic description of galaxy catalogues, including the impact of flux limits and incompleteness and incomplete sky coverage. (4) We extend our treatment of the GMF, and refine it by accounting for the latitude-dependence of the expected deflections \citep{Pshirkov:2013wka}.

The rest of the paper is structured as follows. In Section~\ref{sec:method} we summarise the theory of the AC and XC, as well as describe the UHECR injection models and our treatment of the GMF. Section~\ref{sec:results} collects our results, and Section~\ref{sec:conclusions} concludes with a discussion and an outlook for future work.

\section{Method}\label{sec:method}

\subsection{Theoretical model}\label{ssec:method.thy}

\subsubsection{Sources}

We assume that UHECR sources are numerous (that is, with a comoving density larger than \(10^{-5}\,\text{Mpc}^{-3}\)), steady but faint, so that the probability to observe an event from each source within the life-cycle of a typical detector is much less than one \citep{Waxman:1996hp,Koers:2008ba}.\footnote{In the opposite case UHECR multiplets are expected, see for example \cite{Globus:2022qcr}.} We also assume that their distribution is correlated with the distribution of galaxies detected by a given survey. This can happen in two different ways: either UHECR emitters are a subset of the galaxies in the survey catalogue, or they are a different population of galaxies, which will nonetheless exhibit some correlation with the former ones because they are hosted within the same underlying cosmic large-scale structure.

Each source injects a flux of UHECRs with a power-law spectrum \(\Phi\propto E^{-\gamma}\) up to a very high energy, unless an end-of-spectrum energy cutoff \(\Emax\) is explicitly indicated. Therefore, the integral UHECR flux, (i.e.\ the number of UHECRs detected per unit detector area, time, solid angle and energy) above a certain energy cut \(\Ecut\) (not to be confused with the end-of-spectrum cutoff \(\Emax\)), in a given direction \(\nv\), and for a single species with atomic number \(Z\), reads
\begin{align}\label{eq:cr_flux}
	\Phi(\Ecut,\nv;\gamma,Z)& \deq \frac{\mathcal{E}_0}{4\,\pi}\,\frac{\bar{n}_{\rm s,c}\,\Ecut}{1-\gamma}\left(\frac{\Ecut}{E_0}\right)^{-\gamma} \int_{z_\text{min}}^\infty \de z\, \frac{\alpha(z,\Ecut;\gamma,Z)}{H(z)\,(1+z)}\,\left[1+\delta_{\rm s}(z,\chi\,\nv)\right] \;,
\end{align}
where \(\bar{n}_{\rm s,c}\) is the comoving average number density of sources (which is assumed not to be evolving with redshift, but readily generalised to such a case), \(\mathcal{E}_0\) is the overall emissivity normalisation factor, \(E_0\) the energy at which the flux is normalised, \(\chi\equiv\chi(z)\) is the radial comoving distance to redshift \(z\), \(z_\text{min}\) is the minimum redshift at which we consider UHECR sources (see below), and \(\delta_{\rm s}(z,\chi\,\nv)\) is the UHECR sources density contrast (see also \cite{Urban:2020szk}). The function \(\alpha(z,\Ecut;\gamma,Z)\) is the attenuation function that gives the probability that a UHECR injected with energy above \(\Ecut\) by a source located at redshift \(z\) emitting UHECRs with atomic number \(Z\) and injection slope \(\gamma\) still has \(E>\Ecut\) when it reaches Earth. We calculated the attenuation function by following \(10^6\) simulated UHECRs with \emph{SimProp} v2r4 \citep{Aloisio:2017iyh} with energies above \(E = 10\,\mathrm{EeV}\) (with an upper cutoff of \(\Emax = 10^5\,\mathrm{EeV}\) for \({^{1}}\)H and \({^{16}}\)O, and \(\Emax=280\,\mathrm{EeV}\) for \({^{28}}\)Si), for redshifts \(z\le0.3\). With \emph{SimProp}, we accounted for all energy losses, namely adiabatic losses and losses due to interactions with cosmic microwave background photons and with extra-Galactic background photons according to the model of \citep{Stecker:2005qs}.

We are interested in the anisotropies in the UHECR flux,
\begin{align}\label{eq:cr_ani}
    \Delta(\Ecut,\nv;\gamma,Z) &\deq \frac{\Phi(\Ecut,\nv;\gamma,Z)-\bar{\Phi}(\Ecut;\gamma,Z)}{\bar{\Phi}(\Ecut;\gamma,Z)}\\
    &= \int_{\chi_\text{min}}^\infty\dd\chi\,\phi(\Ecut,\chi;\gamma,Z)\,\delta_{\rm s}(z,\chi\,\nv) \;,
\end{align}
where \(4\,\pi\,\bar{\Phi}(\Ecut;\gamma,Z)\deq \int\dd^2\hat n\,\Phi(\Ecut,\nv;\gamma,Z)\) is the average flux across the sky and \(\chi_\text{min}\) is the comoving distance corresponding to \(z_\text{min}\). In the second line, we have recast the anisotropy as an integral of the UHECR radial kernel
\begin{align}\label{eq:cr_ker}
    \phi(\Ecut,z;\gamma,Z) &= \left[\int_{z_\text{min}}^\infty\dd \tilde{z}\,\frac{\alpha(\tilde{z},\Ecut;\gamma,Z)}{H(\tilde{z})\,(1+\tilde{z})}\right]^{-1} \frac{\alpha(z,\Ecut;\gamma,Z)}{1+z} \;,
\end{align}
where \(H(z)\) is the Hubble parameter and \(\dd\chi/\dd z=1/H\).

\subsubsection{Injection models}

We work with three injection models: \({^{1}}\)H with injection slope \(\gamma=2.6\) at all energies; \({^{16}}\)O with injection slope \(\gamma=2.1\) at all energies; \({^{28}}\)Si with \(\gamma=1.5\) and a sharp \(\Emax=280\,\mathrm{EeV}\) energy cutoff. These are the same three models that were analysed in \cite{diMatteo:2017dtg} and are qualitatively representative of the realistic fits to the observed UHECR energy spectrum.

Heavy nuclei with energy \(E\gg A\,\Ecut\), with \(A\) their mass number, disintegrate rapidly on their way from the source to the Earth into \(A\) nucleons each with energy \(E/A\). For simplicity, and because it is well within our uncertainties (that are driven by the incomplete knowledge of the magnetic fields), we follow \cite{diMatteo:2017dtg} and approximate this process (as well as the decay of neutrons into protons) as instantaneous. In scenarios with no \(\Emax\) as in our \({^{16}}\)O case the result is a component of secondary protons with the same power-law spectrum as the parent nuclei, such as a fraction \(\fpr = A^{2-\gamma}/(A^{2-\gamma}+1)\) from ``light'' elements (which in \emph{SimProp} we count as protons), and the remaining \(1-\fpr\) part from ``heavy'' elements (which in \emph{SimProp} we count as the original nuclei). Notice that we can still use this formula even for \(A=1\), although physically nothing happens in this case. Therefore, in this approximation we substitute
\begin{align}\label{eq:att_p}
    \alpha(z,\Ecut;\gamma,Z) &\rightarrow \fpr\,\alpha(z,\Ecut;\gamma,1) + (1-\fpr)\,\alpha(z,\Ecut;\gamma,Z) \;.
\end{align}
Conversely, because of our choice of maximum energy cutoff at \(\Emax=280\,\mathrm{EeV}\) for \({^{28}}\)Si, all secondary nucleons are produced with \(E\leq10\,\mathrm{EeV}\ll40\,\mathrm{EeV}\leq\Ecut\)), and are hence not included in the analysis. In the \({^{16}}\)O model instead we find \(\fpr\approx0.43\)---this component will be visible in the AC and XC.

\subsubsection{Energy spectrum}

The energy spectrum on Earth, and with it the expected number of UHECR events used to estimate the shot-noise contribution to the UHECR anisotropy map, should be self-consistently derived from the injection spectrum propagated from the source. However, within our level of accuracy, we can more simply choose the final energy spectrum in such a way that it is qualitatively representative of what is observed in current experimental facilities. Therefore, in order to determine the number of UHECR events which in turn fix the levels of Poisson noise in the angular power spectra, we define the differential observed energy spectrum as
\begin{equation}\label{eq:en_spectrum}
    J(E)=
    \begin{cases} 
      J_0\, \left(\dfrac{E}{\text{EeV}}\right)^{-\gamma_1}  & E \leq E_1 \\
      J_0\, \left(\dfrac{E_1}{\text{EeV}}\right)^{-\gamma_1}\,\left(\dfrac{E}{E_1}\right)^{-\gamma_2} & E > E_1
   \end{cases}\;,
\end{equation}
where \(\gamma_1 = 3\), \(\gamma_2 = 5\), \(E_1 = 10^{19.725}\,\mathrm{eV}\), and \(J_0 = 4.28\times10^6/\text{EeV}\) as benchmark values. This choice gives a total number of events above \(\Ecut\), defined as \(\nev\deq \int_{\Ecut}\dd E\,J(E)\), of 1000, 200, 30 events for \(\Ecut=10^{19.6}\,\mathrm{eV}\approx40\,\mathrm{EeV}\), \(\Ecut=10^{19.8}\,\mathrm{eV}\approx63\,\mathrm{EeV}\), \(\Ecut=10^{20}\,\mathrm{eV}=100\,\mathrm{EeV}\), respectively. While these numbers of events are smaller than existing full-sky datasets, see \cite{PierreAuger:2023mvf}, we keep these as baseline values for consistency with our previous analysis in \cite{Urban:2020szk}, but also present results with larger statistics in Section~\ref{sec:results}.

\subsubsection{Galaxies}

The anisotropy in galaxy number counts, using the same language, is defined as
\begin{align}\label{eq:g_ani}
    \Delta_{\rm g}(\nv) &\deq \frac{N_{\rm g}(\nv)-\bar{N}_{\rm g}}{\bar{N}_{\rm g}} =\int_{\chi_\text{min}}^\infty \de\chi \,\phi_{\rm g}(\chi)\,\delta_{\rm g}(z,\chi\,\nv)\;,
\end{align}
where \(N_{\rm g}(\nv)\) is the number of galaxies in a given direction \(\nv\) and \(\bar{N}_{\rm g}\) its average over the celestial sphere. In analogy to the UHECR anisotropies of equation~\ref{eq:cr_ani}, we define the three-dimensional galaxy overdensity as \(\delta_{\rm g}(z,\chi\,\nv)\), and \(\phi_{\rm g}(\chi)\) is kernel of the galaxy distribution, i.e.\ the weighed distribution of galaxy distances (cfr.\ Eq.~\ref{eq:cr_ker}). The galaxy kernel \(\phi_{\rm g}(\chi)\) is given by
\begin{align}\label{eq:g_ker}
	\phi_{\rm g}(\chi)\deq \left[\int \de\tilde\chi \, \tilde\chi^2\,w(\tilde\chi)\,\bar{n}_{\rm g,c}(\tilde\chi)\right]^{-1}\,\chi^2\,w(\chi)\,\bar{n}_{\rm g,c}(\chi) \;,
\end{align}
where \(\bar{n}_{\rm g,c}(\chi)=\chi^{-2}\,\de N_{\rm g}/\de\chi\) is the comoving (volumetric) number density of galaxies in the sample, \(\de N_{\rm g}\) being the angular number density of galaxies in a bin of radial width \(\de\chi\). The quantity \(w(\chi)\) is an optional distance-dependent weight that can be applied to all the objects in the galaxy catalogue, provided their redshifts are known. Assuming Poisson statistics, it can be shown that the optimal weights that maximise the signal of the galaxy-UHECR XC are given by
\begin{align}\label{eq:opt_w}
    w(\chi) &= \frac{\alpha(z,\Ecut;\gamma,Z)}{(1+z)\,\chi^2\,\bar{n}_{\rm g,c}(\chi)} \;.
\end{align}

The properties of the galaxy sample are modelled after the 2MASS Redshift Survey (2MRS) \citep{Huchra:2011ii}, which constitutes one of the most complete full-sky spectroscopic low-redshift surveys, covering the full sky.\footnote{We note here a more recent catalogue \citep{Biteau:2021pru} extending to 350~Mpc, which would be a better choice than the 2MRS catalogue in an actual analysis because of its farther reach and its completeness. Since however we assume an ideal situation here and ``idealise'' the existing 2MRS catalogue, we expect no appreciable difference with the newer catalogue.} After masking out the Galaxy, the 70\% sky coverage for the actual 2MRS catalogue only degrades our signal by a factor of \(\sqrt{0.7}\approx0.84\). For simplicity we also assume that the catalogue is flux-complete, and comment on incompleteness in Section~\ref{ssec:results.cat}. The redshift distribution of galaxies is well fitted by
\begin{align}\label{eq:2MRS}
    \frac{\dd N_{\rm g}}{\dd z}=\frac{N_{\rm g} \,\beta}{\Gamma\left[(m+1)/\beta\right]}\,\frac{z^m}{z_0^{m+1}}\, \exp{\left[-{\left(\frac{z}{z_0}\right)}^\beta\right]}\;,
\end{align}
with \(N_{\rm g}=43\,182\) the total number of sources, \(\beta=1.64\), \(m=1.31\) and \(z_0=0.0266\) \citep{Ando:2017wff}.

Given that the number of detected UHECRs is small (for \(E\gtrsim40\,\mathrm{EeV}\) it is about 1000) we assume that on average each UHECR comes from a different source, which emits UHECRs isotropically, and that all UHECR sources are in the galaxy catalogue. For lack of a better model, we will assume that \(\delta_{\rm g}(z,\chi\,\nv) = \delta_{\rm s}(z,\chi\,\nv)\). This relation may vary by a factor of $O(1)$ if the UHECR flux correlates with other astrophysical properties (e.g.\ optical luminosity, halo mass), as is likely the case.

The fit equation~\ref{eq:2MRS} does not extend to \(z\rightarrow0\) because an actual catalogue will be limited at the distance of the first object. Moreover, the many-sources assumption is valid only above a minimum source density of \(\sim10^{-5}\,\mathrm{Mpc}^{-3}\) \citep{Waxman:1996hp,Koers:2008ba}. This is essentially because, otherwise, a few powerful and nearby sources would outshine the UHECRs flux from more distant ones, and we would observe multiplets of UHECRs from one (or more) directions. Therefore, in all that follows we assume a minimum redshift of \(z_{\rm min}=0.0012\), which roughly corresponds to a minimum distance to the closest galaxy of \(5\,\mathrm{Mpc}\), as is customary in the literature. This is especially important because formally the UHECR kernel equation~\ref{eq:cr_ker} monotonically increases with inverse redshift and peaks at \(z=0\)---indeed, if we did not truncate the galaxy kernel at the minimum \(z_{\rm min}\), we would introduce an unphysical cancellation of cosmic variance (see discussion about figure~\ref{fig:snr_b} in Section~\ref{sec:results}).

\subsubsection{Harmonic spectra}

The two-dimensional fields \(\Delta_a(\nv)\) can be decomposed into their harmonic coefficients
\begin{align}\label{eq:sph_harm}
	\Delta_{\ell m}^a &\deq \int\dd^2\hat n \,Y^*_{\ell m}(\nv)\,\Delta_a(\nv) \;,
\end{align}
where \(Y_{\ell m}\) are the spherical harmonic functions and \(a\in\left\{\text{CR},\,{\rm g}\right\}\) for UHECRs and galaxies, respectively. The covariance of the \(\Delta_{\ell m}\) is the angular power spectrum \(\cS^{ab}_\ell\), defined as
\begin{align}\label{eq:aps}
	\left\langle \Delta^a_{\ell m}\,\Delta^{b\ast}_{\ell'm'}\right\rangle &\deq \delta^{\rm(K)}_{\ell\ell'}\, 
	\delta^{\rm(K)}_{mm'}\,\cS^{ab}_\ell \;,
\end{align}
with \(\delta^{\rm(K)}\) being the Kronecker symbol and the angle brackets stand for the ensemble average.

For broad kernels such as equations~\ref{eq:cr_ker} and~\ref{eq:g_ker} the harmonic, angular power spectrum \(\cS_\ell^{ab}\) between two projected quantities \(\Delta_a\) and \(\Delta_b\) is related to their three-dimensional Fourier-space power spectrum \(P_{ab}(z,k)\) by
\begin{align}\label{eq:cl_limber}
	\cS^{ab}_\ell=\int \frac{\de\chi}{\chi^2} \,\phi_a(\chi)\,\phi_b(\chi)\,P_{ab}\left[k=\frac{\ell+1/2}{\chi},z(\chi)\right] \;,
\end{align}
where \(\phi_a\) and \(\phi_b\) are the radial kernels of both quantities. The synthetic \(P_{ab}(k,z)\) is modelled according to the halo distribution model prescription, adapted to the specifics of the 2MRS catalogue \citep{Peacock:2000qk,Ando:2017wff}.

A given UHECR or galaxy observation consist of both the signal, whose harmonic, angular power spectrum is given by equation~\ref{eq:cl_limber}, and a noise power spectrum \(\cN^{ab}_\ell\), which combined give the observed angular power spectrum:
\begin{align}\label{eq:cell}
    C^{ab}_\ell &\deq \cS^{ab}_\ell+\cN^{ab}_\ell \;.
\end{align}
The noise, being the fields \(\Delta_a(\nv)\) associated to discrete point processes represented by the angular positions of the UHECRs and the galaxies in each sample, is given by
\begin{align}\label{eq:noise}
	\cN^{ab}_\ell=\frac{\bar{N}_{\Omega,a\wedge b}}{\bar{N}_{\Omega,a}\,\bar{N}_{\Omega,b}} \;,
\end{align}
where \(\bar{N}_{\Omega,a}\) and \(\bar{N}_{\Omega,b}\) are respectively the angular number density of points in sample \(a\) and \(b\), and \(\bar{N}_{\Omega,a\wedge b}\) is the angular number density of points shared in common.

For the galaxy clustering auto-correlation, the noise power spectrum reads (see appendix~\ref{app:weighed_noise} for a derivation)
\begin{align}
  \cN^{\rm g\,g}_\ell = \frac{\int \de\chi \,\chi^2\,w^2(\chi)\,\bar{n}_{\rm g,c}(\chi)}{\left[\int \de\chi \, \chi^2\,w(\chi)\,\bar{n}_{\rm g,c}(\chi)\right]^2} \;.\label{eq:noise_gopt}
\end{align}

\subsection{Signal-to-noise ratio}

We estimate the signal-to-noise ratio (\snr) of the UHECR anisotropies under the assumption that the fields being correlated (\(\Delta_{\rm CR}\), \(\Delta_{\rm g}\) in our case) are Gaussian. The explicit expressions for the AC and XC per-\(\ell\) \snr\ read
\begin{align}
	\snr^{\rm CR\,CR}_\ell &= \sqrt{2\,(2\ell+1)\,\left(\frac{\cS^{{\rm CR\,CR}}_\ell}{C^{{\rm CR\,CR}}_\ell}\right)^2} \;,\label{eq:sn_auto}\\
	\snr^{{\rm g\,CR}}_\ell &= \sqrt{(2\ell+1)\,\frac{(\cS^{{\rm g\,CR}}_\ell)^2}{C^{\rm g\,g}_\ell C^{{\rm CR\,CR}}_\ell+(C^{{\rm g\,CR}}_\ell)^2}} \;,\label{eq:sn_cross}
\end{align}
and the total \snr\ is the sum in quadrature, i.e.\
\begin{align}\label{snr_ell}
    \snr^{ab} \deq \sqrt{\sum_{\ell=\ell_{\rm min}}^{\ell_{\rm max}} \left(\snr^{ab}_\ell\right)^2 }\;.
\end{align}
Given the relatively small number of UHECRs currently measured, the shot noise in the UHECR flux is the dominant contribution to the uncertainties, and we are in a situation wherein \(\cS\ll\cN\) for both the AC and the XC. This means that the per-\(\ell\) \snr\ can be approximated as \(\snr^{ab}_\ell \sim \cS^{ab}_\ell / \cN^{ab}_\ell\) to a good accuracy: whereas we retain the full expressions in our computations, this simplified expression guides our analytic intuition when incorporating new effects, such as the magnetic smearing and leaking, and sky cuts.

\subsection{Magnetic fields}\label{ssec:method.mf}

\subsubsection{Magnetic beam}

Upon reaching the Milky Way the UHECRs meet the GMF screen which perturbs their trajectories on their way to the Earth, thus obfuscating the original anisotropy of their arrival direction distribution. The GMF has an amplitude of about a few \(\text{\textmu}\mathrm{G}\), and a complex three-dimensional structure which to date is still poorly understood \citep{Haverkorn:2014jka,Boulanger:2018zrk,Unger:2017kfh,Shaw:2022lqd}. In a simplified treatment we can account for the effects of the GMF by smearing the map of sources below a certain angular scale. The magnetic beam reads
\begin{align}
    {\cal B}(r) &\deq \frac{1}{2\,\pi\,\sigma^2}\exp\left[-\frac{r^2}{2\,\sigma^2}\right] \;, \label{eq:beam_r}\\
    {\cal B}_\ell &\simeq \exp\left[-\frac{\ell\,(\ell+1)\,\sigma^2}{2}\right] \;, \label{eq:beam}
\end{align}
in position and harmonic space, \({\cal B}_\ell\deq\int\dd^2\hat{r}\,{\cal B}(\cos\omega)\,P_\ell(\cos\omega)\), respectively. Here \(r^2=2\,(1-\nv_1\cdot\nv_2)\deq2\,(1-\cos\omega)\) for two directions \(\nv_1\) and  \(\nv_2\), \(P_\ell(\cos\omega)\) is a Legendre polynomial of order \(\ell\) and the width of the Gaussian beam, the displacement \(\sigma\),\footnote{The deflection angle \(\theta\), which is the angle by which a charged particle is deflected from its trajectory in a magnetic field, is not the same as the observable displacement angle \(\sigma\), which is the angle in the sky between the original and actual directions of the charged particle. The two quantities are related as \(\langle\sigma^2\rangle=\langle\theta^2\rangle/3\) \citep{Harari:2002dy}.} is expressed as
\begin{align}\label{eq:smear}
    \sigma(b) \deq \frac{1}{\sqrt2}\left(\frac{40\,\text{EV}}{E/Z}\right) \frac{1\,\text{deg}}{\sin^2 b + 0.15} \;,
\end{align}
where \(b\) is galactic latitude, see \cite{diMatteo:2017dtg,Pshirkov:2013wka} (notice the factor \(1/\sqrt2\) difference with the definition and normalisation of that work)---in the last step of equation~\ref{eq:beam} we have formally assumed that \(\ell\gg1\) but we have verified that within our approximations this can be applied even for \(\ell\sim{\cal O}(1)\).\footnote{The expression equation~\ref{eq:smear} was obtained by mapping the observed variance of rotation measures in the sky to the predicted cosmic-ray deflections, bypassing almost entirely the specifics of the magnetic field model itself. Notice that this model accounts for the small-scale GMF only, and ignores the large-scale component of the GMF, which can be comparable or even larger. Nonetheless, we believe our results will hold in a more realistic model of the GMF for three reasons. (1) The model equation~\ref{eq:smear} is an upper limit on the turbulent deflections from small-scale fields. (2) We always maximise the deflections within a certain area, so we are overestimating the effects of the small-scale GMF. (3) The large-scale GMF acts coherently on the deflections and does not suppresses the global anisotropy in direct proportion to its strength (unlike the small-scale GMF), so its effects on the angular, harmonic correlators are less significant than those of the small-scale GMF.}

In order to be able to treat the problem of magnetic deflections mostly analytically, in this work we smear uniformly across the patch of the sky we are interested in with a displacement angle that is the maximum of equation~\ref{eq:beam} within that patch:
\begin{align}\label{eq:beam_s}
    \cS^{\rm CR\,CR}_\ell \rightarrow {\cal B}_\ell^2\,\cS^{\rm CR\,CR}_\ell \;,~~&~~\cS^{\rm g\,CR}_\ell \rightarrow {\cal B}_\ell\,\cS^{\rm g\,CR}_\ell \;,
\end{align}
for the AC and XC, respectively. Therefore, since the magnetic deflections only depend on \(|b|\), we separate the sky into two regions: a band around the Galactic plane up to a given latitude \(|b|<\bcut\) for which the displacement is \(\sigma(0)\); and the two polar caps \(|b|\geq\bcut\) where we use instead \(\sigma(\bcut)\). We then can add the two \snr\ (once their respective sky fractions have been accounted for) as \(\snr^{ab}_\ell = \sqrt{\left(\snr^{ab,>}_\ell\right)^2 + \left(\snr^{ab,<}_\ell\right)^2}\) where the \(>\) (\(<\)) superscript stands for \(|b|\geq\bcut\) (\(|b|<\bcut\)).\footnote{Using more bands is possible, but the improvement is minimal at the expense of a significant increase in complexity due to the ``leaking'' (see figure~\ref{fig:leak}).} For instance, the maximal displacement for 40~EeV protons are of approximately 4.7~deg when \(b=0\), but if we cut off the galaxy and retain only \(|b|\geq40\,\mathrm{deg}\) we find \(\sigma\approx1.3\,\mathrm{deg}\) in the polar caps covering about 36\% of the sky (for \({^{28}}\)Si we find respectively \(66\,\mathrm{deg}\) and \(18\,\mathrm{deg}\)). Notice that we employ this prescription here simply to have a better estimate of what can be expected from an actual data analysis.

Since the magnetic deflections depend not only on \(Z\) but also on the energy, we can improve on the analysis by binning the UHECR flux in five logarithmic energy bins with logarithmic width of 0.1, and compute the total flux by adding them up proportionally to the UHECR energy spectrum---in the \({^{16}}\)O case we do this for the fractions of oxygen and proton nuclei separately. We then apply to each bin a magnetic smearing obtained from the lowest energy in the bin, that is, \(\Ecuti{i}\), where \(\Ecut\in[\Ecuti{i}\;,\Ecuti{i+1}]\) (for the last bin we set \(\Ecuti{i+1}=\infty\)), as obtained from the energy spectrum equation~\ref{eq:en_spectrum} below. The signal is generalised as
\begin{align}\label{eq:cr_bin}
	\cS^{\rm CR\,CR}_\ell& \rightarrow \sum_{ij} Q_i\, Q_j \, {\cal B}_{\ell,i}\, {\cal B}_{\ell,j}\,\cS^{\rm CR\,CR}_\ell \;,\\
	\cS^{\rm g\,CR}_\ell& \rightarrow \sum_i Q_i \,{\cal B}_{\ell,i}\,\cS^{\rm g\,CR}_\ell \;,
\end{align}
where \(Q_i\) are the fractions of flux in the energy bin \(i\) and \({\cal B}_{\ell,i} \deq {\cal B}_\ell[\sigma(\Ecuti{i})]\).

\subsubsection{Magnetic leaking}

Because in our treatment the magnetic smearing equation~\ref{eq:smear} is latitude-dependent, in Section~\ref{sec:results} we will assess the performance of the AC and XC in the case we split the sky in two in order to contain the effects of the smearing. Physically, smearing the map of sources with a certain angle reproduces the effects of the deflections experienced by UHECRs while propagating through the GMF. Therefore, in the vicinity of \(|b|=\bcut\) it is possible that some UHECRs ``leak'' between the two regions.

\begin{figure}
\centering
    \includegraphics[width=\textwidth]{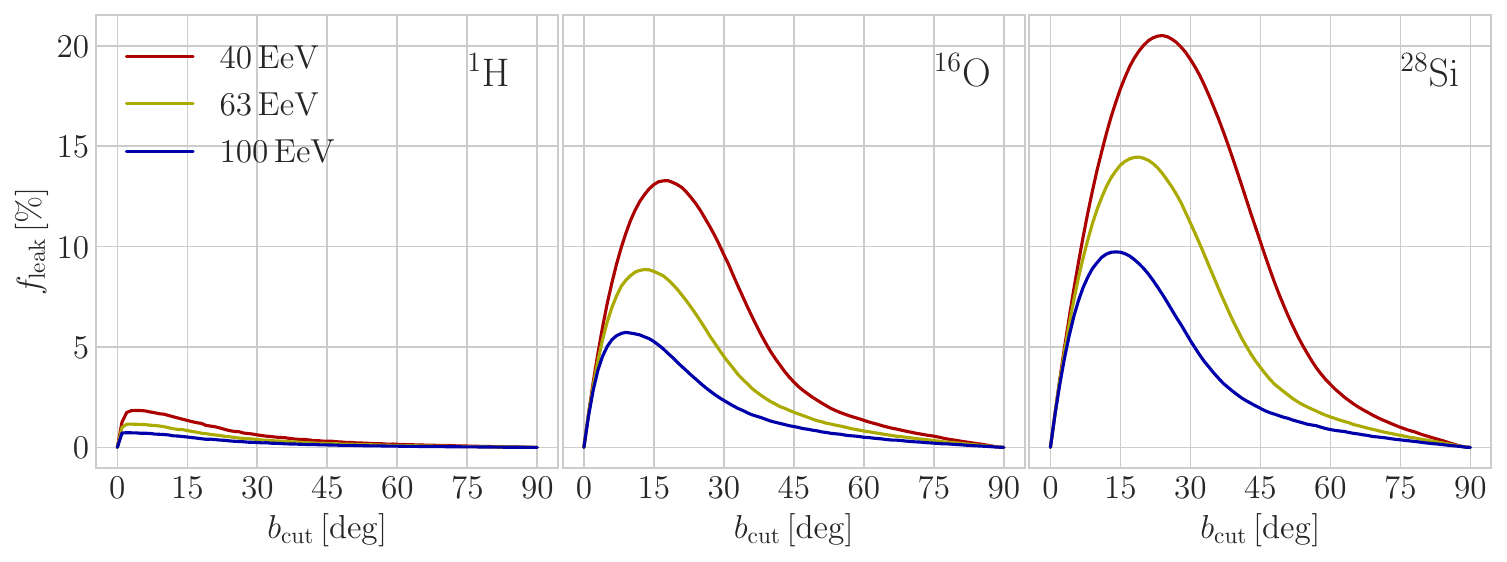}
	\caption{Percentage of synthetic UHECRs in the polar caps that leaked out from the inner, high-deflection region (around the Galactic plane) for \({^{1}}\)H (left), \({^{16}}\)O (centre) and \({^{28}}\)Si (right).}
	\label{fig:leak}
\end{figure}

In order to assess how many UHECRs do so we have simulated the propagation of \(10^6\) UHECRs uniformly distributed on the sphere in a random magnetic field that follows the latitude-dependence spelt out in equation~\ref{eq:smear}; the cell size was chosen to be 10~pc, and each UHECR is propagated until it leaves the sphere which has a radius of 10~kpc (making on average approximately few\(\times10^3\) steps). The overall strength of the magnetic field is normalised to \(1.4\,\text{\textmu}\mathrm{G}\) in order to reproduce a distribution of displacements that follows equation~\ref{eq:smear} for 95\% of the UHECRs: this is conservative in the sense that if we require that \emph{all} the UHECRs stay below the upper limit equation~\ref{eq:smear} we would have on average smaller deflections. Note that there are some small deviations from this distribution at very large deflections above \(50\,\mathrm{deg}\): in that case the actual distribution of displacements is slightly wider with \(b\); hence, the choice of keeping the same formula equation~\ref{eq:smear} even in this case is conservative. For each \(\bcut\), we then count what is the fraction of UHECRs that originally was in the inner, high-displacement region and finished in the polar caps. This fraction of UHECRs would have to be assigned a larger smearing angle than \(\sigma(\bcut)\); to be conservative, we count the leaked fraction of UHECRs as isotropic, which means that they only contribute to the noise and not the signal. What we find is that even for the lowest energies \(\Ecut=40\,\mathrm{EeV}\) and heavier elements (\({^{28}}\)Si with \(Z=14\)), the maximum contamination from the inner, high-deflection regions never exceeds 20\% (at \(\bcut\approx25\,\mathrm{deg}\)), while it is below 10\% for \(\Ecut=100\,\mathrm{EeV}\) and irrelevant for \({^{1}}\)H in all cases---for a cut at \(\bcut=40\,\mathrm{deg}\) we have, for the worst case of \({^{28}}\)Si, less than 15\%, 7\%, 4\% for \(\Ecut=40\,\mathrm{EeV}\), \(\Ecut=63\,\mathrm{EeV}\) and \(\Ecut=100\,\mathrm{EeV}\), respectively. This is shown in figure~\ref{fig:leak}. Moreover, since we bin the UHECR energy spectrum in energy bins, the amount of leaked UHECRs is in fact even smaller---notice that we neglect the small fluctuations between energy bins here, as they are subdominant.

\section{Results}\label{sec:results}

\begin{figure}
\centering
    \includegraphics[width=\textwidth]{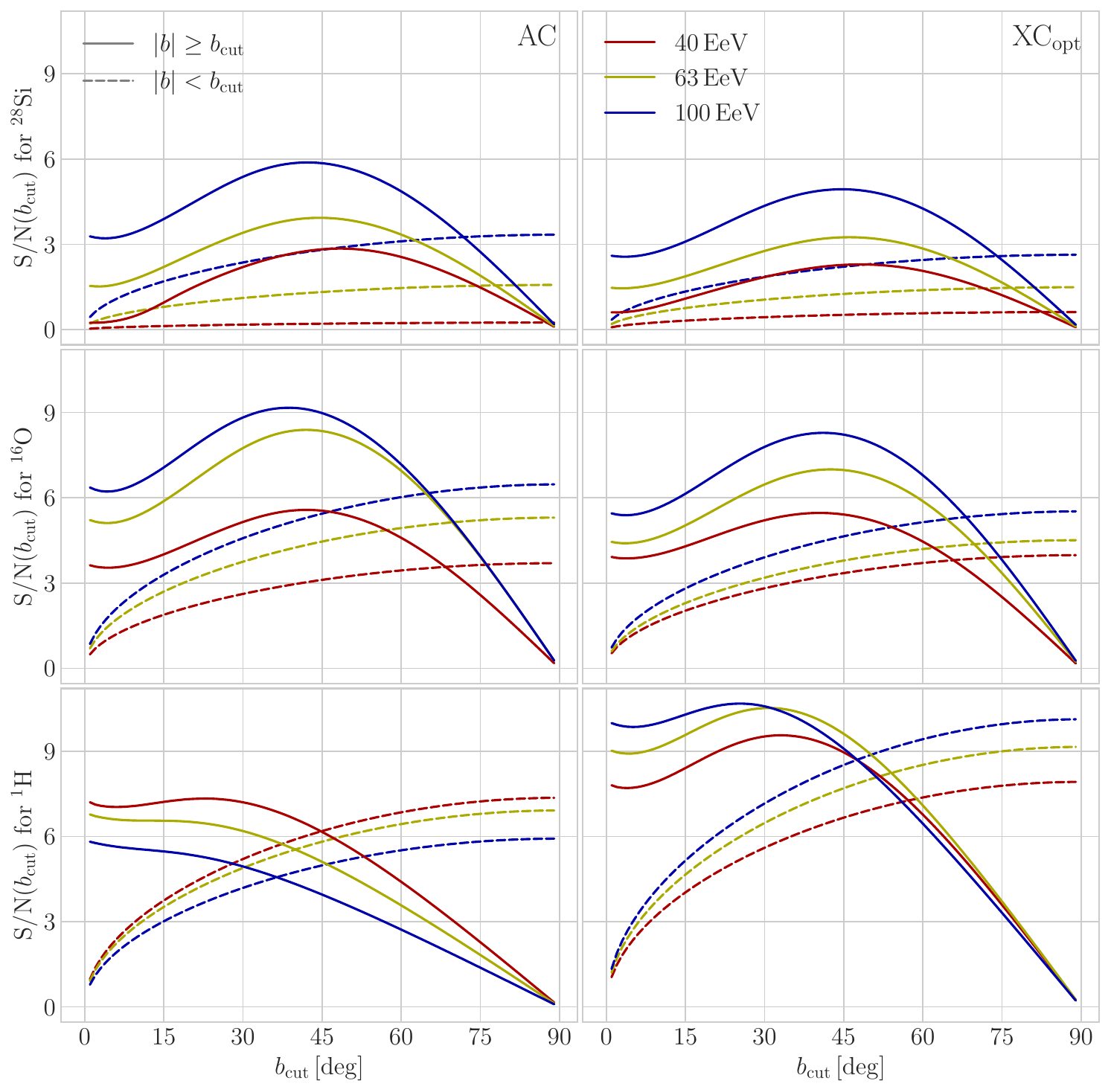}
	\caption{Total \(\snr\) for \(\ell=[2,1000]\) as a function of the cut \(\bcut\) for a north-south-symmetric split for which the sky fraction in the polar caps is \(\fsky=1-\sin\bcut\) for \({^{1}}\)H injection (bottom panels), \({^{16}}\)O injection (middle panels) and \({^{28}}\)Si injection (top panels). The left column shows the AC and the right column shows the XC (optimised). The solid lines refer to the polar caps with \(|b|\geq\bcut=40\,\text{deg}\). The dashed lines refer to the complementary region around the Galactic plane.}
	\label{fig:snrtot_b}
\end{figure}

The magnetic deflections decrease as we look away from the Milky Way. Within this model there is therefore a trade-off: on the one hand choosing a region in the sky where the displacements are small improves the signal because magnetic smearing suppresses angular anisotropies in proportion to the displacement angle \(\sigma\); on the other hand the smaller the fraction of the sky \(\fsky\) we observe, the larger cosmic variance. In order to quantify this trade-off, in figure~\ref{fig:snrtot_b} we show the total \(\snr\) for AC and optimal XC in the range \(\ell=[2,1000]\) for the three injection models \({^{1}}\)H, \({^{16}}\)O and \({^{28}}\)Si, each of them for the three energy cuts of \(\Ecut\approx40\,\mathrm{EeV}\), \(\Ecut\approx63\,\mathrm{EeV}\) and \(\Ecut=100\,\mathrm{EeV}\), where we divide the sky in two regions: the polar caps with \(|b|\geq\bcut\) and for which the displacement is given by \(\sigma(\bcut)\) are shown in solid, whereas the \snr\ from the inner region around the Galaxy (with displacement \(\sigma(0)\)) is shown as a dashed line.\footnote{In this section we always only consider the optimal XC as the \snr\ of the non-optimised XC is always much smaller.} The most notable feature here is that, if UHECRs are heavy nuclei, the \snr\ in the polar caps peaks at around \(\bcut=40\,\mathrm{deg}\), corresponding to having about \(35\%\) of the sky in the caps and \(65\%\) around the Galactic plane. This \snr\ is larger than the full-sky \snr\ with maximal deflections. In UHECRs are protons, separating the sky in a low-deflection and a high-deflection region does not bring about much improvement---this can be understood because heavy elements are more affected by the magnetic deflections; nonetheless, when adding in quadrature the \snr\ from the two regions, the split sky gives better results. We will therefore use \(\bcut=40\,\mathrm{deg}\) in what follows.

\begin{figure}
\centering
    \includegraphics[width=\textwidth]{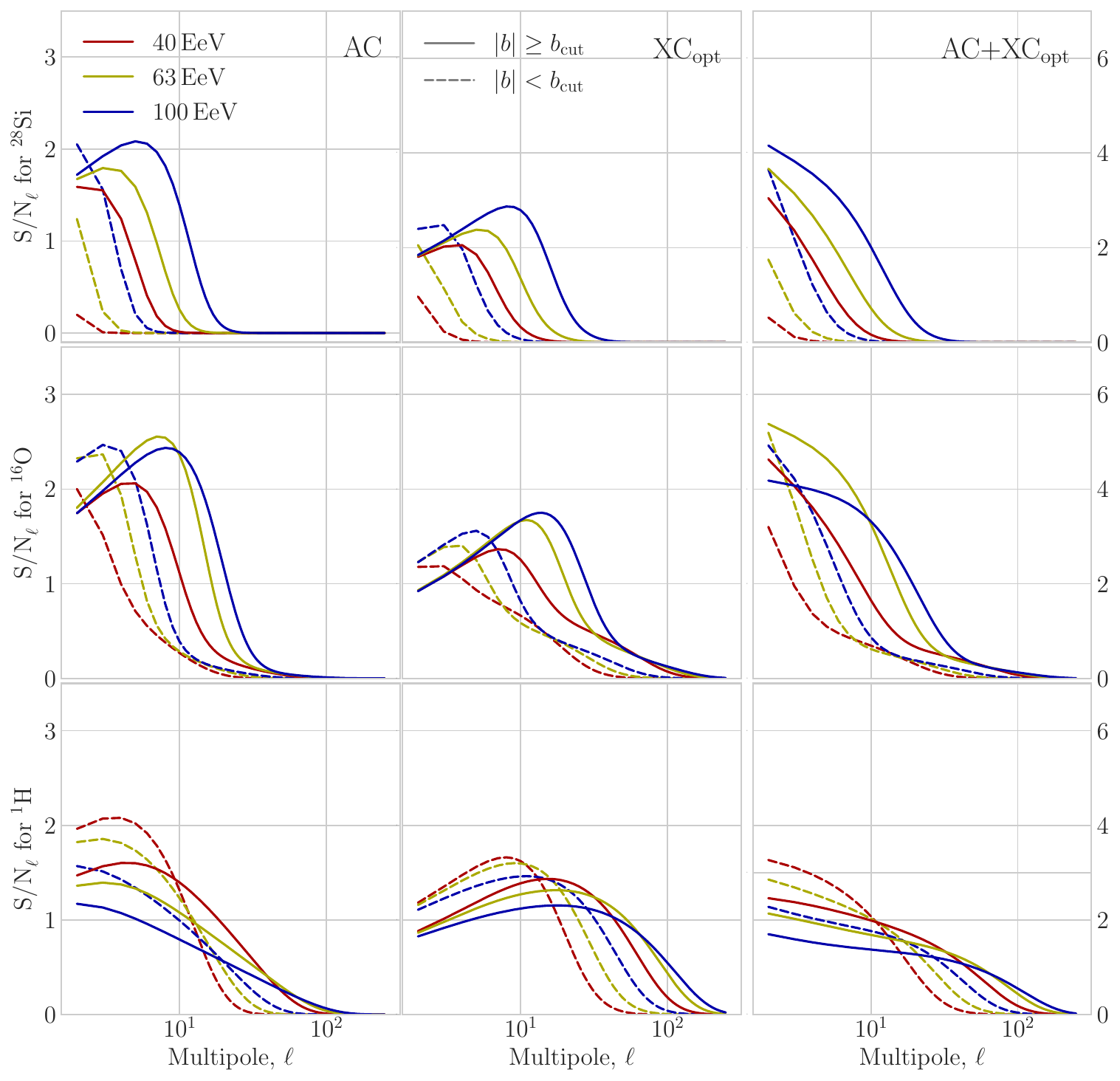}
	\caption{\(\snr_\ell\) for \({^{1}}\)H, \({^{16}}\)O, \({^{28}}\)Si for a North-South symmetric mask with \(\bcut=40\,\deg\), for the same energy cuts as in figure~\ref{fig:snrtot_b}. The left column is the AC, the middle column is the XC (optimised) and the right column is the AC+XC (optimised).}
	\label{fig:snr_b}
\end{figure}

In figure~\ref{fig:snr_b} we show the expected per-\(\ell\) \snr\ for the three injection models \({^{1}}\)H, \({^{16}}\)O and \({^{28}}\)Si. The most important feature is the shift of the sensitivity between the AC and the XC, as the latter is more sensitive to smaller angular scales, i.e.\ higher multipoles \(\ell\). This is in line with the expectations from our previous work \citep{Urban:2020szk}, and it holds up for all the three injections models we consider, as well as after applying the GMF smearing---which, however, severely suppresses the anisotropy, especially at those small scales. We also observe how, as expected from figure~\ref{fig:snrtot_b}, the difference between polar caps and Galactic plane when \(\bcut=40\,\mathrm{deg}\) is much more pronounced fo \({^{28}}\)Si and \({^{16}}\)O than for \({^{1}}\)H, the latter being much less sensitive to the GMF.

The total anisotropic power for both the AC and the XC (if optimal weights are applied) is detectable for all injection models and energy cuts, as is shown in figure~\ref{fig:snrtot_b}. However, unless we use the \emph{combined} power of the AC and the XC, individual multipoles are below the level of detection (which we quantify at \(3\,\sigma\), approximately \(\snr_\ell\approx3\)). Such combined analysis (rightmost panels, dubbed `AC+XC') is worth being discussed for another, peculiar feature. Looking at figure~\ref{fig:snr_b}, we observe that the AC+XC shows an overall enhancement of \(\snr_\ell\) for all injection models and energy cuts particularly on the largest angular scales, i.e.\ low multipoles, even as the AC and XC separately may drop as \(\ell\rightarrow1\)---notice the different y-axes ranges in the rightmost panels compared to the AC and XC alone. This is caused by a cosmic-variance mitigation effect that is known to occur when multiple tracers of the cosmic large-scale structure are employed simultaneously \citep[][see also \citealp{2014MNRAS.442.2511F} for a first application to harmonic-space power spectra]{2009PhRvL.102b1302S}. What happens is that, in order to analyse the AC and XC jointly, we have to take into account the \textit{cross-covariance} between the two signals \citep[see][Section~3.1]{Urban:2020szk}. It is important to notice that both signals trace the same, underlying large-scale cosmic structure. Therefore, when analysing both at the same time, the inverse covariance in the joint \(\snr\) involves ratios of \(C^{\rm CR\,CR}\) and \(C^{\rm g\,CR}\), in which the common, stochastic contribution from the matter density distribution effectively cancels out. While cosmic-variance cancellation happens at all scales, this effect is much more pronounced at low \(\ell\) because cosmic variance scales as \(1/(2\ell+1)\) so it is larger (hence, it pollutes the signal more) at low \(\ell\); moreover, cosmic variance cancellation is effective when sample variance dominates over shot noise, which is increasingly true as we move towards low \(\ell\). This face raises the exciting possibility that some low multipoles, e.g.\ the quadrupole \(\ell=2\), could be within reach of current data already, if the combined AC and XC were used. We leave a full discussion of this possibility to future work.

\begin{figure}
\centering
    \includegraphics[width=\textwidth]{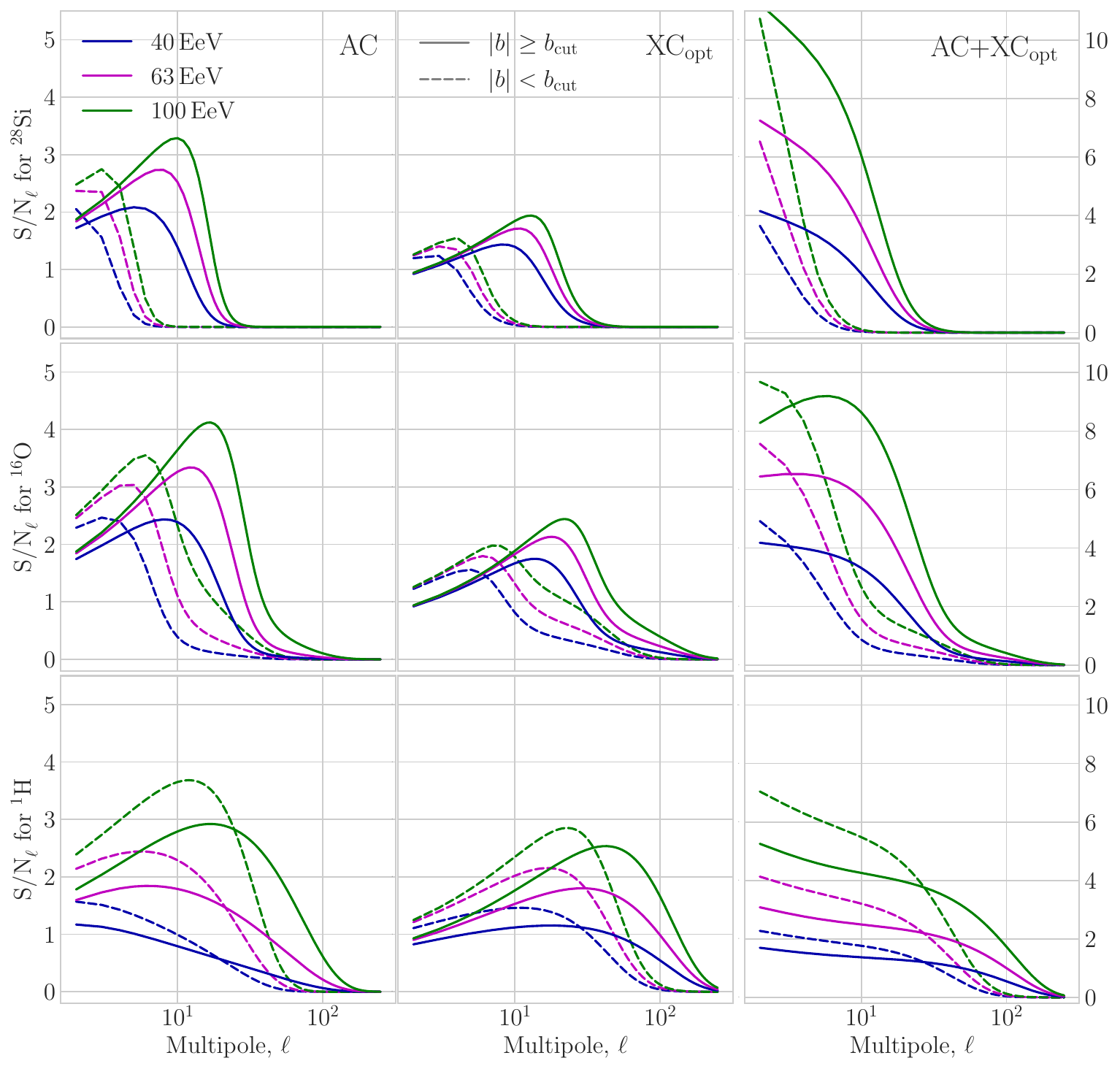}
	\caption{\(\snr_\ell\) for \({^{1}}\)H, \({^{16}}\)O, \({^{28}}\)Si for a north-south-symmetric mask with \(\bcut=40\,\deg\), for \(\Ecut=100\,\mathrm{EeV}\) with \(\nev=30\) (blue, same as in figure~\ref{fig:snrtot_b}), \(\nev=100\) (magenta) and \(\nev=300\) (green).}
	\label{fig:snr_b_nev}
\end{figure}

In order to assess how a future facility with access to larger exposure could improve on this result, in figure~\ref{fig:snr_b_nev} we compare the \(\Ecut=100\,\mathrm{EeV}\) results of  figure~\ref{fig:snr_b} with synthetic UHECR data sets containing approximately three and ten times more events, namely \(\nev = 100\) and \(\nev = 300\) events. We note how the AC improves much faster than XC, as expected given the fact that the AC entirely relies on \(\nev\) to trace the underlying distribution of sources (the noise for the AC goes as \(1/\nev\)), see equations~\ref{eq:sn_auto} and~\ref{eq:sn_cross}.

\subsection{Catalogues}\label{ssec:results.cat}

\paragraph{Incomplete sky coverage} Galaxy surveys are never able to observe the whole sky. In the simplest case, a large region around the Galactic plane must be masked to avoid systematics from dust extinction and star contamination. In addition to this, ground-based observatories never have access to the whole celestial sphere, and observing conditions also place constraints on the final survey footprint. From the point of view of data analysis, a limited sky area has a number of consequences. First of all, not all angular scales will be accessible, with the minimum multipole being approximately \(\ell_{\rm min}\sim\pi/(2\,\fsky)\). Secondly, a smaller sky area will impact how noisy the signal will be. An accurate estimate of this effect depends on the details of the sky mask, but for \(\fsky\gtrsim\mathcal O(10^{-1})\) we can account for this  by multiplying the (Gaussian) error bars on the power spectrum by a factor \(1/\sqrt{\fsky}\). In addition to this, an incomplete sky coverage effectively corresponds to applying a binary mask \(v(\hat{\bm n})\) to the underlying target field \(f(\hat{\bm n})\)---binary in the sense that  \(v\) is either \(1\) if the corresponding pixel has been observed, or \(0\) if not. However, the simple product \(v(\hat{\bm n})\,f(\hat{\bm n})\) in real space becomes a convolution in harmonic space. This fact has two major consequences: \(i)\) the na\"ive estimator for the harmonic-space power spectrum becomes biassed and such a bias has to be corrected for; and \(ii)\) different \(\ell\)-modes become coupled because of the presence of the mask, and the data covariance is no longer diagonal in \(\ell\)-\(\ell^\prime\), even on linear scales. Nonetheless, from the point of view of the results presented here, this last point has no major impact \citep{2011MNRAS.414..329C}, since we are dealing with very large sky areas, for which the effects of the mask will be small.

Furthermore, it is worth emphasising that a limited sky area for a galaxy survey will also impact the cross-correlation between galaxies and UHECRs, since the areas of the sky observed by the two experiments will not in principle be the same. As a matter of fact, some degree of correlation is expected even in non-overlapping areas, since both galaxies and UHECR emitters trace the same underlying cosmic large-scale structure. However, to be conservative, we only consider the intersection of the sky patches observed by the two experiments.

\paragraph{Flux limitations} The main impact of flux incompleteness in the galaxy catalogue is the potential for some of the detected UHECRs being sourced by galaxies not included in the catalogue. The quantitative impact of this on the results presented here would be relatively mild as long as the galaxy catalogue is able to cover the same physical volume over which most of the UHECR sources are distributed. Although the cross-correlation noise bias would change (see Eq.~\ref{eq:noise}), the main noise contribution to the power spectrum covariance comes from the noise in the auto-correlations, which would stay the same. On sufficiently large scales (which are the only ones where the cross-correlation can be measured effectively, anyway), the amplitude of the cross-correlation is controlled by the spatial overlap between both tracers (i.e.\ their ability to trace the same matter fluctuations), regardless of their overlap at the level of individual sources. Thus, as a rough estimate, if, say, 2\% of UHECR sources are missing from the catalogue this will translate into a maximum of 2\% loss of XC power, because of the 2\% smaller galaxy kernel.

This picture would change if the UHECR sources were located at redshifts that are significantly larger than those probed by the galaxy catalogue, but this is unlikely given the redshift dependence of the attenuation factor for typical cosmic rays, all the more so for UHECRs.

\paragraph{Other catalogues} To date, a rather large amount of galaxy surveys for cosmology has been conducted, resulting into a likewise large availability of galaxy catalogues to employ for cross-correlation studies. Moreover, future observational campaigns plan to observe deeper (in both redshift and magnitude), meaning that in the next future even more data will be available.

This notwithstanding, to our aim of learning about UHECRs via cross-correlations with galaxy clustering, we are limited to a small volume around the Earth, for the radial kernel of UHECRs decreases rapidly with increasing redshift. This means that we need a catalogue with as many as possible \textit{local} galaxies, so that the two signals correlate (technically, this is so to maximise the support of the integral in Eq.~\ref{eq:cl_limber}). For this reason, we have focussed on 2MRS, which is the largest catalogue available at low redshift, both in terms of number density and of sky coverage.

However, it is worth keeping in mind that galaxy surveys differ from one another not only for what concerns redshift range, observed area, and total number of detected objects, but also because they target different galaxy populations. Such galaxy populations will have, for instance, different bias, which may make the cross-correlation signal stronger. Moreover, their astrophysical properties will also differ, making them better targets for their rate of emission of UHECRs. Therefore, for an actual data analysis, it would be interesting to compare UHECR cross-correlations among various catalogues, as they may teach us something about the properties of UHECR sources.

\section{Conclusions and outlook}\label{sec:conclusions}

In this work we have studied the angular, harmonic cross-correlation between the distribution of galaxies in the local Universe (our synthetic galaxy catalogue peaks at redshift \(z\approx0.03\)) and a full-sky set of UHECRs in numbers comparable to what has been detected with current facilities. We have considered three injection models (\({^{1}}\)H, \({^{16}}\)O, \({^{28}}\)Si) which bracket realistic fits to the observed UHECR spectrum and composition, and three energy cuts at \(40\,\mathrm{EeV}\), \(63\,\mathrm{EeV}\) and \(100\,\mathrm{EeV}\). Lastly, we have included the effects of intervening magnetic fields using a data-driven toy-model for smearing.

We find that, with statistics comparable with current data, both the AC and XC for individual multipoles will be rather weak and remain below the threshotd for detection, mostly because of the magnetic deflections. However, if we combine the AC and the XC in the most optimal statistics for the detection of the anisotropy, we are able reach detection levels of \(3\sigma\) and above in the \({^{16}}\)O and \({^{28}}\)Si cases at nearly all energies considered here for very large scales (low multipoles). This is thanks to a cancellation of cosmic variance that arises when employing multiple tracers of the same underlying distribution (in our case the large-scale structure). Cosmic variance cancellation motivates further studies, aiming at improving on the searches for anisotropy at low multipoles, e.g., the quadrupole \(\ell=2\) (see~\cite{TelescopeArray:2023waz}). In some cases, a marginal detection at \(3\sigma\) for individual, optimised XC multipoles is within reach with ten times more events than currently available, as would be the case for instance with the next generation of detectors such as POEMMA or GRAND. Lastly, the total power \(C^{ab}\deq\sum_\ell C^{ab}_\ell\) is detectable with SNRs well above~5 for the AC and XC (optimised) for both \({^{1}}\)H and \({^{16}}\)O at all energies, with peaks of \(\snr\approx10\) for the XC for \({^{1}}\)H at \(100\,\mathrm{EeV}\) and \(\snr\approx8\) for the AC and XC of \({^{16}}\)O at most energies. \({^{28}}\)Si is the most impacted by the magnetic deflections, as expected, but still has \(\snr\approx6\) for the AC and \(\snr\approx5\) for the optimised XC (cutting the sky at \(\bcut=40\deg\)). We remark that all our results, except for the simulated energy losses of propagating UHECRs, are analytic; this enables us to quickly asses some properties of the harmonic observables in different scenarios.

Our findings can serve as further motivation to search for the XC (and the AC+XC) with existing data from the Telescope Array (located in the Northern Hemisphere) and the Pierre Auger Observatory (located in the Southern Hemisphere), following the lead of the AC analyses performed by the Auger-TA joint working group on arrival directions~\citep{TelescopeArray:2023waz}. Another application is the use of the angular, harmonic power spectra to discriminate between different UHECR injection models, in particular to help determine the UHECR composition, see for example~\cite{Tanidis:2022jox}.

For the future we plan to move away from the analytic analysis we have kept to thus far and perform fully-fledged propagation simulations of UHECRs through realistic models of the GMF. This will enable us to interpret the possible detection of harmonic correlations from the data. This is especially needed for the promising AC+XC quadrupole, which we predict could already be detectable with existing data.

\acknowledgments
FU thanks Armando di Matteo for help with \emph{SimProp}. SC acknowledges support from the Italian Ministry of University and Research (\textsc{mur}), PRIN 2022 `EXSKALIBUR – Euclid-Cross-SKA: Likelihood Inference Building for Universe's Research', and from the European Union -- Next Generation EU. DA acknowledges support from the Beecroft Trust, and from the Science and Technology Facilities Council through an Ernest Rutherford Fellowship, grant reference ST/P004474/1.

\appendix
\counterwithin{figure}{section}

\section{Exposures}\label{app:expo}

The exposures of the Telescope Array Collaboration (TA) and Pierre Auger Observatory (PAO) experiments are strongly \(b\)-dependent. This means that masking certain regions of the sky changes not only cosmic variance, as quantified by \(\fsky\), but also the density of UHECR events within the unmasked region \(\bar{N}_{\Omega,{\rm CR}}\rightarrow\bar{N}_{\Omega,{\rm CR}}(\bcut) \deq N_{\rm CR}/[4\,\pi\,\fsky(\bcut)]\), and with it the noise.

\begin{figure}
\centering
    \includegraphics[width=0.75\textwidth]{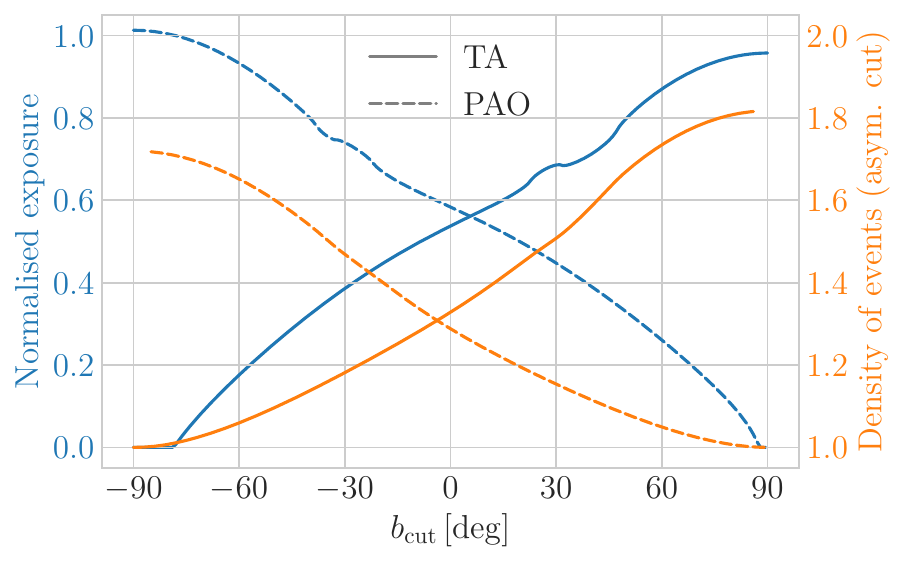}
	\caption{Left y-axis (blue colour): normalised exposures \(\omega(b)\) averaged over \(l\) for TA (solid) and PAO (dashed). Right y-axis (orange colour): \(\bar{N}_{\Omega,{\rm CR}}(\bcut)\), normalised to 1 for a full sky, when only the north, resp.\ south, polar cap beyond \(|\bcut|\) are unmasked.}
	\label{fig:expos}
\end{figure}

Exposures above \(10\,\mathrm{EeV}\) are purely geometric and, save for very subdominant sidereal effects, only depend on declination \(\delta\) as
\begin{align}
    \omega & \propto \cos\lambda\,\cos\delta\,\sin\alpha_m + \alpha_m\,\sin\lambda\,\sin\delta \;,
\end{align}
where \(\lambda\) is the latitude of the detector (for TA \(\lambda=39.2\,\deg\) and for Auger \(\lambda=-35.2\,\deg\)), and the parameter \(\alpha_m\) is
\begin{align}
    \alpha_m &\deq
    \begin{cases}
        0 & \xi>1,\\
        \pi & \xi<-1,\\
        \arccos\xi & \,\mathrm{otherwise}
    \end{cases}\;,
\end{align}
with
\begin{align}
    \xi &\deq \frac{\cos\theta_z - \sin\lambda\,\sin\delta}{\cos\lambda\,\cos\delta} \;,
\end{align}
and \(\theta_z\) the maximum zenith angle (we use \(\theta_z = 55\,\deg\) for TA and \(\theta_z = 60\,\deg\) for Auger) \citep{Sommers:2000us}. The exposure in Galactic coordinates \((l,b)\) can be computed thanks to \(\sin\delta = \sin\delta_\text{NGP}\,\sin{b} + \cos\delta_\text{NGP}\,\cos{b}\,\cos(l_\text{NCP}-l)\) where the north Galactic pole has \(\delta_\text{NGP} = 27.13\,\deg\) and the north celestial pole has \(l_\text{NCP} = 122.9\,\deg\) in the J2000 system---here we can also use \(\cos\delta = \sqrt{1-\sin^2\delta}\) because \(\delta\in[-\pi/2,\pi/2]\). Because the GMF deflections equation~\ref{eq:smear} do not depend on \(l\) we average over it: \(\omega(b) \deq \int\de l\,\omega(b,l)/(2\,\pi)\). In figure~\ref{fig:expos}, left panel, we show the exposures of TA and Auger normalised such that \(\int\de b\sin{b}\,\omega(b)=1\). In this approximation the exposure \(\omega(b)\) rises nearly linearly with \(b\); thence, if we apply a north-south-symmetric mask the density \(\bar{N}_{\Omega,{\rm CR}}(\bcut)\) is nearly independent of \(\bcut\) (the increase in density in one hemisphere being compensated by the loss in the opposite hemisphere) and, in practice, we recover the case of uniform exposure.

An alternative approach would be to cut the sky from \(b=-\pi/2\) to \(b=\bcut\) and keep everything north of that for TA (or any north-located experiment), and cut the sky from \(b=\pi/2\) to \(b=\bcut\) and keep everything south of that for Auger (or any south-located experiment). We show the UHECR event densities with this choice of cuts in figure~\ref{fig:expos}, right panel. In this case the expression for the deflection \(\sigma(b)\) is
\begin{align}\label{eq:smear_asym}
    \sigma = \frac{1}{\sqrt2}\left(\frac{40\,\text{EV}}{E/Z}\right) \frac{1\,\text{deg}}{\Theta(\pm b)\,\sin^2b+0.15} \;,
\end{align}
where \(\Theta\) is the Heaviside unit step function and the sign \(+\) is for TA and the \(-\) is for Auger. We have verified that for all cases under consideration (\({^{1}}\)H, \({^{16}}\)O, \({^{28}}\)Si) it is best to keep the full sky. If the magnetic deflections are larger, for example when we consider heavier primaries, lower energies, or stronger magnetic fields, it might however be advantageous to keep only the north (south) polar cap for TA (Auger).

\section{The galaxy noise power spectrum}
\label{app:weighed_noise}
Let us start by considering the measurement of galaxy number count fluctuations in real space and focus on a purely Poisson process. The (weighed) number of galaxies at angular position \(\nv\) is
\begin{equation}
    N_{\rm g}^w(\nv)=\int\de\chi \,N_{\rm g}(\nv,\chi)\,w(\chi)\;,
    \label{eq:Ng_mean}
\end{equation}
where \(N_{\rm g}(\nv,\chi)\,\de\chi\) is the number of galaxies in direction \(\nv\) in the radial-comoving distance interval \([\chi,\chi+\de\chi]\), and \(w(\chi)\) is an arbitrary weight assigned to it (\(w\equiv1\) being the unweighted case). Then, we have
\begin{align}
    \left\langle\left[N_{\rm g}^w(\nv)\right]^2\right\rangle&=\int\de\chi \,\int\de\chi'\,w(\chi)\,w(\chi')\,\left\langle N_{\rm g}(\nv,\chi)\,N_{\rm g}(\nv,\chi')\right\rangle\nonumber\\
    &=\int\de\chi\,\int\de\chi' \,w(\chi)\,w(\chi')\,\left[\left\langle N_{\rm g}(\nv,\chi)\right\rangle\,\langle N_{\rm g}(\nv,\chi')\rangle+\left\langle N_{\rm g}(\nv,\chi)\right\rangle\,\delta^{\rm(D)}(\chi-\chi')\right]\nonumber\\
    &=\left[\int\de\chi\,w(\chi)\,\bar N_{\rm g}(\chi)\right]^2+\int\de\chi\,w^2(\chi)\,\bar N_{\rm g}(\chi)
    \;,
\end{align}
with \(\delta^{\rm(D)}\) denoting the Dirac distribution and \(\bar N_{\rm g}(\chi)\) being the mean number of galaxies per unit comoving distance in \([\chi,\chi+\de\chi]\). In the second line we have used the fact that: \textit{i)} number counts are Poisson processes, for which the variance is equal to the mean; and \textit{ii)} there is no correlation unless the radial position is the same. Thence,
\begin{equation}
    {\rm Var}\left[N_{\rm g}^w(\nv)\right]=\int\de\chi\,w^2(\chi)\,\bar N_{\rm g}(\chi)\;.
    \label{eq:Ng_variance}
\end{equation}

Now, \(\Delta_{\rm g}\) is defined in equation~\ref{eq:g_ani} as the number density \textit{contrast}, i.e.\ the fluctuations in number counts relative to the mean. Therefore, by dividing equation~\ref{eq:Ng_variance} by equation~\ref{eq:Ng_mean} squared and taking the average over all possible directions in the sky, we eventually obtain the Poissonian shot noise for the harmonic-space power spectrum of a weighed galaxy distribution, viz.\ equation~\ref{eq:noise_gopt}.

\bibliography{references}

\end{document}